\documentclass[12pt]{article}
\usepackage{amsmath}
\usepackage{graphicx}
\usepackage{float}
\newcommand{\be}{\begin{equation}}
\newcommand{\ee}{\end{equation}}
\newcommand{\ba}{\begin{eqnarray}}
\newcommand{\ea}{\end{eqnarray}}

\begin{document}
\title{{\bf Possible Superluminal Propagation\\ inside Conscious Beings}\!\!
\thanks{Alberta-Thy-1-20, arXiv:2001.11331 [physics.gen-ph]}}

\author{
Don N. Page\!\!
\thanks{Internet address:
profdonpage@gmail.com}
\\
Theoretical Physics Institute\\
Department of Physics\\
4-183 CCIS\\
University of Alberta\\
Edmonton, Alberta T6G 2E1\\
Canada
}

\date{2020 January 31}

\maketitle
\large
\begin{abstract}
\baselineskip 20 pt

The fact that first-person conscious perceptions or sentient experiences have many bits of information strongly suggests that they are produced nonlocally by the effects of many atoms, say by nonlocal quantum operators.  If these nonlocal operators act back on the quantum state of the atoms, they could lead to evolution in which signals propagate superluminally, violating the usual causality of local quantum field theory.  Although there is not yet any direct evidence that nonlocal operators associated with psycho-physical parallelism act back on the quantum state, it is not totally implausible that this might be the case.  In principle the resulting superluminal propagation might be observable by sending signals across brain regions (neural correlates of consciousness) that lead to conscious perceptions.\\

\end{abstract}
\normalsize
\baselineskip 17 pt
\newpage

\section{Introduction}

It is not yet known the full relationship between the present level of a physics description of the universe (most fundamentally in terms of quantum physics) and a description of consciousness as a first-person sentient experience (not consciousness in the third-person sense of a certain type of information processing in the brain, though that often accompanies the first-person sense).  Perhaps the simplest framework for part of this relationship is that that sentient experiences come with different measures (analogous to probabilities, but without true randomness) that are the expectation values of positive operators, one for each possible sentient experience \cite{Page:1995dc,Page:1995kw,Page:2001ba,Page:2011sr,Page:2017mpl}, which I call {\it awareness operators}.  The details of these quantum awareness operators are not yet known, so this is a framework for the relationship between consciousness and quantum physics rather than a detailed theory for the relationship or psycho-physical parallelism.  I have called this framework {\it sensible quantum mechanics} in technical papers \cite{Page:1995dc,Page:1995kw,Page:2011sr,Page:2017mpl} and {\it mindless sensationalism} in a paper for those outside physics \cite{Page:2001ba}; here I shall use the former phrase or its acronym SQM.\\

In the simplest version of SQM, the conscious perceptions and the corresponding awareness operators have no effect on the quantum state, so that the conscious perceptions and their measures are epiphenomena, being determined by the quantum state and the awareness operators, but not having any back reaction on the quantum state, which in the Heisenberg picture is a constant quantum state, fixed once and for all.  (Certainly in SQM, when a conscious perception occurs, the quantum state is not collapsed to any eigenstate of the corresponding awareness operator, or to any other state different from what that Heisenberg state is at all times, such as the state one would get by acting on the original Heisenberg state by the awareness operator.  It is just that the expectation value of the awareness operator in the fixed Heisenberg state gives the measure for the conscious perception, but this measure occurs in the space of conscious perceptions, which is not to be identified with the space of quantum states.)\\

However, it is conceivable that in an approximation to the quantum gravity state of the universe in which there is a time variable, in the Schr\"{o}dinger picture the quantum state evolves with a Hamiltonian that includes a contribution from the awareness operators.  One might still say that the conscious perceptions themselves remain epiphenomena, not affecting the quantum state, but then the awareness operators themselves, which, along with the quantum state, determine the measures of the conscious perceptions, also influence the time evolution of the quantum state in the Schr\"{o}dinger picture.  I personally do not think it is {\it a priori} very probable that the awareness operators do contribute to the Hamiltonian, since to me that would seem to complicate the theory, but since, as we shall see, the consequences of this contribution could introduce a radically new element into physics, I do think it is well worth investigating whether or not this is the case.\\

The radical new element that the contribution of the awareness operators to the Hamiltonian seem likely to make is to make the evolution of the quantum state {\it acausal}, in the sense that they may lead to the possibility of signals propagating faster than the usual speed limit $c$ of the universe, the ``speed of light.''  This arises because it seems most plausible that the awareness operators are {\it essentially nonlocal}, by which I mean not the integrals of local operators as the usual Hamiltonian in local quantum field theory is.  In this way the physics of consciousness might lead to superluminal propagation of signals.\\

The argument that the awareness operators (and hence the Hamiltonian, {\it if} the awareness operators do contribute to the Hamiltonian) are essentially nonlocal is that the number of bits of information in a conscious perception seems to be much higher than the number of fundamental local fields that are excited in conscious beings such as ourselves.  Inside humans, there are significant excitations of the graviton, photon, neutrino, electron, quark, and gluon fields, but not much else that we have found experimentally.  Surely the number of bits of information in an alert human conscious perception is much higher than the number of such local fields that are not near their ground states (though see \cite{Norretranders}), whereas if the awareness operators were just integrals over space of awareness operator densities that are simple combinations of these local fields at each point, there would not be enough of them to give the very many different conscious perceptions that seem to be possible from the apparent amount of information in the conscious perceptions that we experience.  (There is here the assumption that the awareness operator densities do not involve high powers of the fundamental boson fields or high derivatives of any of the fundamental fields, which would be unlike what seems to be the case for the Hamiltonian density for the usual Hamiltonians considered in local quantum field theories such as the Standard Model.  Such high powers and/or high derivatives in the awareness operator densities conceivably could lead to a large enough number of awareness operators to explain the not-too-low apparent information in a typical alert human conscious perception, though here I shall assume that this is not the case.)\\

On the other hand, if the awareness operators are integrals over space or spacetime of nonlocal awareness operator densities, even without high powers or high derivatives of the fields, there could be a large enough number of awareness operators to explain the fairly large apparent amount of information in our alert human conscious perceptions.  Then if the Hamiltonian density included nonlocal awareness operator densities, the Schr\"{o}dinger-equation evolution of the quantum state would include these nonlocal contributions and could lead to superluminal propagation of signals, faster than the ``speed of light'' $c$ that is the speed limit for local quantum field theory.  Since the awareness operator densities would only have significant contributions inside conscious beings, one would expect the superluminal propagation also only to be significant inside conscious beings, which would be consistent with the fact that no convincing superluminal propagation has been seen in previous searches that, so far as I know, have all been done outside conscious beings.\\

\section{Magnitudes of possible superluminal propagations}

It would be extremely exciting to be able to observe possible superluminal propagation inside conscious beings, even though I would guess that the chance that this actually occurs is low.  It is an effect that has not been searched for with the sensitivity plausibly required.  Unfortunately, even if this superluminal propagation does occur, it is plausible that it is so small and/or so weak that it would be very difficult to observe, so I am afraid that I am not at all confident that it can be confirmed or refuted within my lifetime, or even within the lifetime of my younger colleagues.\\

To illustrate how difficult it might be, I shall give some extremely crude guesses, but there probably are other ways to give better estimates of the magnitudes of the effect.  For these estimates, I shall use Gaussian natural units with $\hbar = c = 4\pi\epsilon_0 = k_{\mathrm Boltzmann} = 1$, but here not using Planck units in which $G = 1$, since I am assuming that the awareness operators for human conscious perceptions do not depend significantly upon gravitational interactions.  I shall also use $m_e$ for the mass of the electron, which in Gaussian natural units is equivalent to $9.109\,383\,7015(28)\times 10^{-31}$ kg, or to $8.187\,103\,7769(25)\times 10^{-14}$ J, or to $7.763\,440\,7063(23)\times 10^{20}$ s$^{-1}$, or to $2.589\,605\,0748(8)\times 10^{12}$ m$^{-1}$ in various SI units, and the dimensionless fine-structure constant $\alpha = e^2 = e^2/(4\pi\epsilon_0\hbar c) = 0.007\,297\,352\,5693(11) = 1/137.035\,999\,084(21)$.  The mass of the proton, $m_p$, is $1836.152\,673\,43(11) \sim 2^2 3^3 (2^3+3^2)$ times the mass of the electron, $1.672\,621\,923\,69(51)\times 10^{-27}$ kg $=1.503\,277\,615\,98(46)\times  10^{-10}$ J $=1.425\,486\,240\,79(44)\times 10^{24}$ s$^{-1} = 4.754\,910\,2812(15)\times 10^{15}$ m$^{-1}$.\\

Let us assume that the the number of bits in an alert human conscious perception is roughly those of the relevant fields in $N^3$ water molecules of mass $\sim 18\, m_p$ and volume $3\times 10^{-29}$ m$^3 = 200\, a_0^3$, where $a_0 = 1/(m_e\alpha) = 5.291\,772\,109\,03(80)\times 10^{-11}$ m is the Bohr radius.  Then a cube of $N^3$ water molecules would have an edge length of about $L = 5.8\, N a_0 \sim 3\,N\times 10^{-10}$ m, so this is a rough estimate of the minimum linear nonlocality size of an awareness operator density for an alert human conscious perception.  It is even more uncertain what $N$ is.  From my own awareness of visual images, I would conjecture that $N^3$ is at least several hundred, so I might guess that a minimum value for $N$ is of the order of 10, which would give $L = L_1 \sim 3$ nm (three nanometers).  Light would take $\Delta t_1 = L_1/c \sim 10^{-17}$ s = 10 attoseconds to cross this region, but if the awareness operator density has a nonlocality of this size and contributed suitably to the Hamiltonian density, signals might be able to propagate across this region instantaneously (in the frame of the awareness operator density with the dominant expectation value, which would plausibly be close to that of the matter of the conscious being, e.g., the brain of the human having the conscious perception whose measure is given by the expectation value of that awareness operator).\\

On the other hand, the region within a human brain that contributes to each conscious perception might be considerably larger, leading to an acausality of the awareness operator density of the same length scale, $L \gg 5.8\, N a_0$, say up to the linear size $L = L_2 \sim 0.1$ m of the parietal, temporal, and occipital lobes of the cerebral cortex that contains the posterior cortical hot zone that is a candidate \cite{KMBT} for the full neural correlates of consciousness (NCC), the minimal region where a full conscious perceptions is produced, corresponding in SQM to the region of nonlocality of the awareness operator density, where a significant contribution occurs to the expectation value of the awareness operator and hence for the measure of that conscious perception.  If so, the time reduction for the superluminal propagation could be increased up to $\Delta t_2 = L_2/c \sim 3\times 10^{-10}$ s (three hundred picoseconds).  Therefore, the signal propagation time reduction from the contribution of the awareness operator density to the Hamiltonian density might be expected to be in the range from 10 attoseconds to 300 picoseconds.\\

Besides the question of the spatial range of the possible superluminal propagation and the corresponding timescale, there is also the question of the energy contributed to the Hamiltonian by the nonlocal awareness operator density.  This is even more uncertain than the spatial range and superluminal propagation time reduction.  There is no logical reason that I know of that the energy contribution could not be arbitrarily small or even zero, which I do regard as the simplest and most plausible possibility, since there is no requirement that I know of that the awareness operators (whose expectation values are conjectured to give the measures of the corresponding conscious perception) should have any effect on the quantum state.  However, under the exciting possibility that the awareness operators do make a contribution, one would like to make at least some crude guesses as to what energies they might contribute.\\

One of the lowest possibly plausible positive energy estimates for the contribution of the awareness operators to the Hamiltonian would be the excitation energy of a proton or neutron in a box of size $L$, which with momentum $p \sim \hbar/L$ would be $E_1 \sim p^2/m_p \sim \hbar^2/(m_p L^2)$, which for $L$ varying from $L_1 \sim 3\times 10^{-9}$ m to $L_2 \sim 0.1$ m runs from $E_{11} \sim \hbar^2/(m_p L_1^2) \sim 10^{-24}$ J $\sim 10^{-5}$ eV (ten micro-electron volts) down to $E_{12} \sim \hbar^2/(m_p L_2^2) \sim 10^{-39}$ J $\sim 10^{-20}$ eV (ten zepto-electron volts).  Assuming that electrons are more probably the particles directly involved in the awareness operators, a more plausible set of estimates for the energy would be $E_2 \sim \hbar^2/(m_e L^2)$, which for $L$ varying from $L_1 \sim 3\times 10^{-9}$ m to $L_2 \sim 0.1$ m runs from $E_{21} \sim \hbar^2/(m_e L_1^2) \sim 10^{-21}$ J $\sim 10^{-2}$ eV (ten milli-electron volts) down to $E_{22} \sim \hbar^2/(m_e L_2^2) \sim 10^{-36}$ J $\sim 10^{-17}$ eV (ten atto-electron volts), about three orders of magnitude greater than the estimates $E_1$ from the excitation energy of a proton.  A third perhaps equally plausible set of estimates would be that it is the electromagnetic field that gives the dominant contribution to the awareness operators and contributes an energy $E_3 \sim \hbar c/L$ , which for $L$ varying from $L_1 \sim 3\times 10^{-9}$ m to $L_2 \sim 0.1$ m runs from $E_{31} \sim \hbar c/L_1 \sim 10^{-16}$ J $\sim 10^3$ eV (one kilo-electron volt) down to $E_{32} \sim \hbar c/L_2 \sim 3\times 10^{-24}$ J $\sim 2\times 10^{-5}$ eV (20 micro-electron volts).  Thus we get energy estimates that range from $10^{-20}$ eV to $10^3$ eV, but with no guarantee that the perturbation of the Hamiltonian by the putative effect of the awareness operators is within this range.  Indeed, most plausibly, there would be no effect, but since a nonlocal effect would be so exciting, it seems well worth considering the nonzero possibilities for the energy perturbation in the possible nonlocal part of the Hamiltonian density.\\

\section{How to Search for Possible Superluminal Propagation}

Finding possible superluminal propagation in the physics of consciousness would be astounding new knowledge about our universe, but searching would also be extremely challenging for the following three reasons:\\

(1)  The superluminal propagation would only be expected to arise, if at all, where the awareness operator densities are significant, that is, inside conscious beings (and inside whatever parts of those beings are directly responsible for their consciousness, the full neural correlates of consciousness, such as some parts of their brains).  However, very few fundamental physics experiments are performed inside conscious beings, since it is much easier performing physics experiments on inanimate matter.\\

There is also the ethical issue that one should not cause significant pain or other damage for conscious beings (though I would be willing to suffer the small pain of a needle injection or a mild headache if that were the cost of an experiment on myself, or more pain if I could be convinced that the probability of finding superluminal propagation were much higher than I pessimistically suspect it is).  Therefore, one would want to use a minimally invasive procedure.  Whether or not inserting thin probes into the brain of a conscious being is sufficiently noninvasive to keep any pain (both before and after the experiment) very small, I do not know, but this would be worth further investigation.  Another possible procedure might be to focus x-rays onto one side of a region believed to produce consciousness and then look for x-rays coming out from the opposite side before the original incident x-rays come out from that side, though then one would need to keep the incident x-ray intensity at a safe level.\\

(2)  Since the region in a brain directly producing consciousness (where the awareness operator density is significant) is likely to be small, and perhaps extremely small if it just needs to contain enough atoms to give the number of bits in a conscious perception, the very small time needed for a signal at the speed of light to traverse that region is likely to make it very difficult to detect a superluminal signal propagating faster (e.g., effectively instantaneously in the frame of the brain).\\

(3)  The possibly very low energies plausibly associated with the acausal corrections to the Hamiltonian density from the putative contributions from the awareness operator densities give at least two challenges:  (a)  When the energies are similar to $\hbar/\Delta t = \hbar c/L$, as $E_3 \sim \hbar c/L$ is, or much less, as $E_1 \sim \hbar^2/(m_p L^2)$ and $E_2 \sim \hbar^2/(m_e L^2)$ are, the reduction in the propagation time, $\Delta t \sim L/c$, would be buried in the fluctuations from the uncertainty principle, so one would need a lot of data to give a large enough signal-to-noise ratio to become confident of a truly superluminal signal.  (b)  If the signals propagating out from the part of the brain where the awareness operator densities are high enough that they might produce superluminal propagation (if indeed they contribute to the Hamiltonian density) are to have high enough frequencies that they can propagate through the rest of the brain essentially at the speed of light (e.g., at x-ray frequencies), the quanta of such signals are likely to be of much higher energy than the mean energy contribution to the Hamiltonian by the awareness operators.  This suggests that such signals are likely to be exponentially suppressed.  This problem with looking for x-ray signals coming out sooner than would be possible causally might require the insertion of probes into the two sides of the part of the brain directly responsible for conscious perceptions, which would be much more invasive and liable to cause pain and/or damage.\\

Therefore, there is a great challenge to be able to design an experiment that is ethical in not causing undue pain or damage to any conscious being, that is able to resolve the very small timescales of the plausibly possible superluminal propagation, and that is able to record a statistically significant signal despite the very small energies expected for the possible superluminal propagation effect.\\

\section{Conclusions}

The apparent nonlocality of the neural correlates of consciousness and hence of the densities of the awareness operators (whose expectation values are postulated to give the measures of the corresponding conscious perceptions) suggests the possibility of superluminal propagation of signals within conscious beings, {\it if} the awareness operator densities also contribute to the Hamiltonian density and hence to the evolution of the quantum state (unlike the probably simpler hypothesis that nothing related to consciousness acts back on the quantum state).  Although this possibility is perhaps improbable, if it were true, it would be such an astounding surprise that it seems well worth investigating.  Unfortunately, the prospect of actually being able to detect such superluminal propagation, even if it does exist, appears to be so challenging that a definitive test does not to my na\"{\i}ve eyes look very likely in the foreseeable future.  However, the challenge could be highly worth the effort.  

\section{Acknowledgments}

This work has been supported in part by the Natural Sciences and Engineering Council of Canada. Many of the people whom I have remembered as being influential in my formulation of my ideas on consciousness are listed at the end of \cite{Page:1995dc}.  This particular paper was also motivated by a critical reading of Sean Carroll's stimulating book \cite{Carroll}, which (though without itself pointing this out) helped me to realize that we have virtually no direct evidence whether or not the physics of conscious beings is for all practical purposes fully described by what Carroll calls, following Frank Wilczek, the Core Theory (the standard model of particle physics plus Einstein's general relativity theory of gravity).

\end{document}